\documentclass[referee]{raa}            

\usepackage{graphicx,times}             
\usepackage{natbib}
\usepackage{amssymb,amsmath}
\bibpunct{(}{)}{;}{a}{}{,}
\usepackage{ctex} 
\usepackage{array}

\newcolumntype{C}[1]{>{\centering\arraybackslash}p{#1}}
\renewcommand{\abstractname}{Abstract}
\renewcommand\tablename{Table}
\renewcommand\figurename{Figure}

      \newcommand{\ps}{\,{\rm s}$^{-1}$}
    
    \newcommand{\km}{\,{\rm km}}

\newcommand{\HII}{H~{\sc ii}}
       
\newcommand{\twCO}{$^{12}$CO}  \newcommand{\thCO}{$^{13}$CO}
\newcommand{\CeiO}{C$^{18}$O}

\begin{document}

  \title{A Comparative Study of TeV Gamma-Ray Sources with Various Objects}

   \volnopage{Vol.0 (202x) No.0, 000--000}      
   \setcounter{page}{1}          

   \author{Xin Zhou 
      \inst{1, 2,*}\footnotetext{$*$Corresponding Authors.}
   \and Ji Yang
      \inst{1, 2}
   \and Yang, Su
      \inst{1, 2}
   \and Xuepeng, Chen
      \inst{1, 2}
   \and Yang, Chen
      \inst{3,4}
   \and Yan, Sun
      \inst{1, 2}
   \and Qing-Zeng, Yan
      \inst{1}
   \and Shaobo, Zhang
      \inst{1, 2}
   }
   \institute{Purple Mountain Observatory, Chinese Academy of Sciences, 10 Yuanhua Road, Nanjing 210023, People's Republic of China; {\it xinzhou@pmo.ac.cn}\\
	\and State Key Laboratory of Radio Astronomy and Technology, Purple Mountain Observatory, Chinese Academy of Sciences, 10 Yuanhua Road, Nanjing 210023, China;
        \and School of Astronomy \& Space Science, Nanjing University, 163 Xianlin Avenue, Nanjing 210023, People's Republic of China;
	\and Key Laboratory of Modern Astronomy and Astrophysics, Nanjing University, Ministry of Education, Nanjing 210023, People's Republic of China
   }

\abstract{
We investigate the relationships between LHAASO TeV gamma-ray sources and various kinds of objects, including pulsar wind nebulae (PWNe), supernova remnants (SNRs), \HII\ regions, microquasars, and OB associations. 
We propose a Randomization-Adjusted Overlap Correlation (RAOC) method to statistically assess association probabilities and evaluate association proportions across catalogs.
The results reveal statistically significant overlaps between LHAASO sources and SNRs, PWNe, and microquasars, supporting their role as important contributors to TeV gamma-ray emission. 
The estimated association proportions of LHAASO sources are 0.19$\pm$0.08 with SNRs, 0.20$\pm$0.04 with PWNe, and 0.027$\pm$0.008 with microquasars.
The proportion of the gamma-ray sources associated with the subsample of shell-type SNRs is $\sim$0.1.
While \HII\ regions also show potential association, particularly with the KM2A component, their large self-overlap ratio complicates precise estimation. In contrast, OB associations exhibit a high probability of chance coincidence, suggesting their limited contribution to TeV gamma-ray emission.
Our analysis of TeV gamma-ray emission capabilities shows that $\sim$60\% of PWNe are gamma-ray bright in both the WCDA and KM2A energy ranges.
For SNRs and microquasars, the TeV gamma-ray bright fraction is $\sim$10\%. 
The subsample of PWNe associated with molecular clouds (MCs) shows enhanced gamma-ray emission. 
Furthermore, positional analysis reveals a systematic offset of the gamma-ray sources overlapping with PWNe toward the associated MCs. These findings imply a role for MCs in PWN gamma-ray production.
Additionally, self-correlation analysis indicates that about 70\% of the WCDA and KM2A gamma-ray components share a common origin. The study also identifies selection effects in existing SNR catalogs and notes clustering among approximately 30\% of \HII\ regions within larger star-forming regions. 
Further multi-wavelength studies are needed to elucidate the physical mechanisms underlying these associations.
}

   \authorrunning{X. Zhou, et al.}            
   \titlerunning{The origin of TeV gamma-ray sources}  

   \maketitle

%
%
\section{Introduction}           
\label{sec:intro}
The origin of Galactic cosmic rays (CRs) remains one of the most fundamental problems in astrophysics. 
CRs are mainly charged particles that are deflected by interstellar magnetic fields, hence, the gamma-rays produced by their interactions are essential for tracing their sources \citep{Blasi2013}. 
Among the potential accelerators, supernova remnants (SNRs) and pulsar wind nebulae (PWNe) are considered the most promising accelerators of Galactic CRs \citep{Blasi2013, Cao+2024a, Cao+2024, Cao+2025a}. 
PWNe are efficient at producing leptonic CRs, however, their contribution to the hadronic component of CRs is still unknown. 
Other kinds of sources, such as \HII\ regions \citep{Hampton+2016} and black hole-jet systems \citep{Cao+2025}, have also been discovered to be associated with gamma-ray emission and are considered to be important contributors to Galactic CRs.

Recent advances in gamma-ray astronomy, powered by high-sensitivity observations from instruments like the Large High Altitude Air Shower Observatory (LHAASO), have revealed a large population of high-energy gamma-ray sources. Many of these sources are spatially coincident with known SNRs and PWNe \citep{Cao+2024}. However, distinguishing physical associations from chance projections in the crowded Galactic plane is challenging.

Traditional approaches to source association often involve detailed multi-wavelength studies of individual objects. For instance, the gamma-ray properties of SNRs Kes 73, G35.6$-$0.4, and G51.26+0.11 have been studied in detail using multi-wavelength data, confirming the role of Molecular Clouds (MCs) in enhancing gamma-ray emission \citep{Liu+2017, Zhang+2022, Zhong+2023}.
While these case studies provide deep physical insights, a statistical analysis of the associations between the LHAASO TeV source catalog and various potential counterparts is lacking. 
MCs associated with the CR accelerators can be illuminated by energetic protons, further complicating the situation \citep[e.g.,][]{MitchellCelli2024}.
Comprehensive statistical studies are essential for disentangling the contributions of different astrophysical objects to the observed gamma-ray emissions.


In this study, we conduct a statistical analysis of the associations between LHAASO gamma-ray sources and various kinds of objects, including PWNe, SNRs, and \HII\ regions. We propose a Randomization-Adjusted Overlap Correlation (RAOC) method to assess the statistical significance of the spatial overlap and to evaluate the proportion of associated sources. Additionally, we particularly focus on understanding how the presence of MCs influences the gamma-ray detectability of SNRs and PWNe.
The observational data and the RAOC method are described in Sections~\ref{sec:data} and \ref{sec:method}, respectively. The results are presented in Section~\ref{sec:result}, including the association statistics between LHAASO sources and different kinds of objects in Section~\ref{sec:asso}, and a specialized study of PWN-MC association candidates in Section~\ref{sec:pwnmc}. Section~\ref{sec:sum} summarizes our conclusions.


\section{Data}\label{sec:data}
The CO data were obtained from the Milky Way Imaging Scroll Painting (MWISP\footnote{http://www.radioast.nsdc.cn/mwisp.php}) project. The MWISP project was carried out using the Purple Mountain Observatory Delingha 13.7~m millimeter wavelength telescope, which is an unbiased survey of \twCO/\thCO/\CeiO\ (J=1--0) emission lines \citep[see][for details]{Yang+2026}.
The velocity resolutions are 0.17~\km\ps for \twCO\ and 0.16 \km\ps\ for the other two lines.
The pointing and tracking accuracy was better than 5$''$. The data were gridded to 30$''$ spacing, with a half-power beamwidth of $\sim$51$''$. 
After linear baseline subtraction, typical rms noise levels are $\sim$0.45~K per channel for \twCO\ and $\sim$0.24~K for \thCO\ and \CeiO. 

The TeV gamma-ray sources are taken from the first Large High Altitude Air Shower Observatory (LHAASO) catalog \citep{Cao+2024}.
The catalog includes sources detected by the Water Cherenkov Detector Array (WCDA) and the Kilometer Squared Array (KM2A) array, which cover the energy ranges 1--25 TeV and above 25 TeV, respectively. 
The size of each gamma-ray source is taken as the larger of its extension and its position uncertainty \citep[i.e., $\sigma_{p,95,stat}$ and $r_{39}$ in Table~1 of][]{Cao+2024}, to account for all possible contributions.
The SNR and pure PWN samples within the MWISP coverage ($1^\circ < l < 230^\circ$ and $|b| < 5.5^\circ$) are obtained from the \cite{Green2019} catalog\footnote{http://www.mrao.cam.ac.uk/surveys/snrs/snrs.info.html} and the high-energy SNR catalog SNRcat\footnote{http://www.physics.umanitoba.ca/snr/SNRcat} \citep{FerrandSafi-Harb2012}. 
The pure PWNe are excluded from the SNR sample.
The \HII\ region sample is introduced from one of the most complete catalogs of \HII\ regions in the Galaxy: the Wide-field Infrared Survey Explorer catalog of Galactic \HII\ regions \citep{Anderson+2014}.
Microquasars positions are obtained from \cite{MirabelRodriguez1999} and \cite{Remillard+2006}, each assigned a nominal size of $10"$.
OB associations are taken from \cite{Chemel+2022}, where the size is defined as the angular diameter enclosing 68\% of the member stars.

%
%

\section{Method} \label{sec:method}
\begin{figure}
\centering
\includegraphics[width=15cm, angle=0]{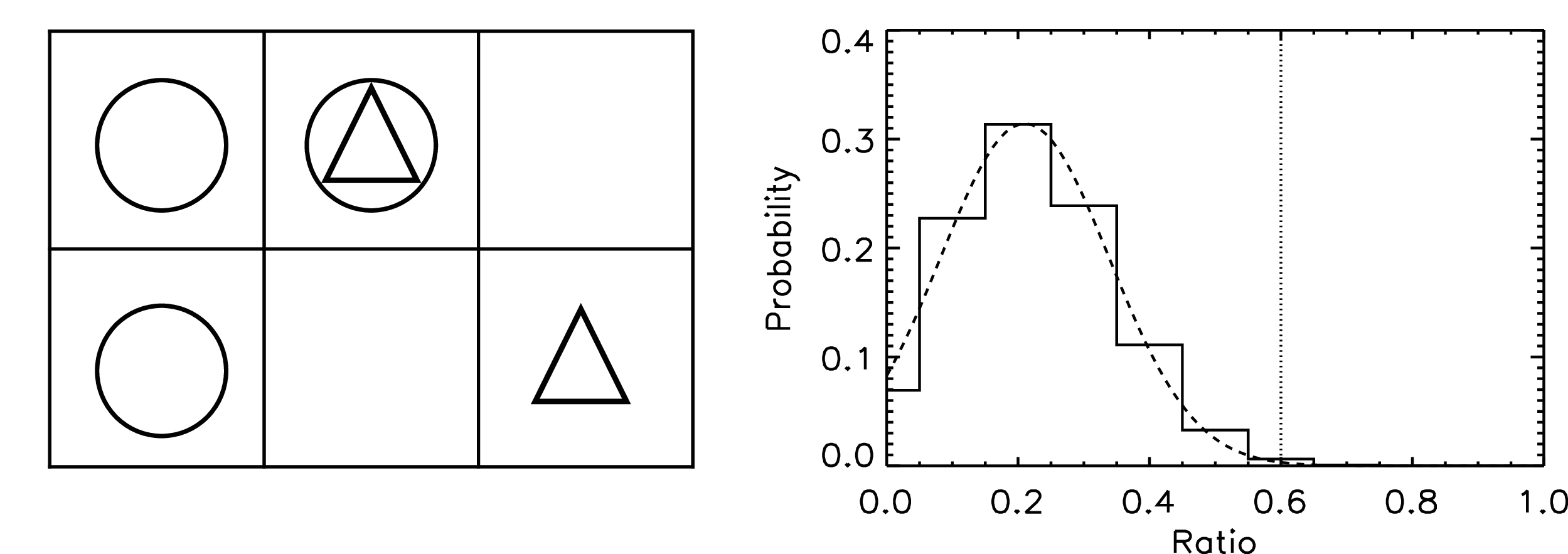}
\caption{Left: Example case of two types of samples (two triangles and three circles) distributed in six grids. Right: Distribution of the proportion of overlapping items for a toy model. The toy model has two types of samples that are randomly distributed in 100 grids. The two samples contain 20 and 30 elements, respectively. Ten elements are predetermined to overlap. The histogram shows the distribution of the proportion of randomly overlapping items in the first type of sample, excluding the pre-selected overlapping elements.
The ratio distribution is fitted with a Gaussian function, and the fitting result is shown as a dashed line. The dotted line shows the most probable final overlap ratio, taking into account both randomly overlapping and pre-selected overlapping elements.
}
\label{f:toy}
\end{figure}
The overlap between two sets of samples distributed within a region may be due to a physical correlation or to coincidence.
The proportion of chance overlap can be examined by comparing them to randomly distributed samples. For simplicity, Figure~\ref{f:toy} shows a toy model. In this model, two types of samples are randomly distributed over a grid; for instance, there are $n_1$ and $n_2$ samples in $n$ grids. The probability of observing $i$ overlaps is $P_i=C_{n-n_2}^{n_1-i} C_{n_2}^{i}/C_{n}^{n_1}=C_{n-n_1}^{n_2-i} C_{n_1}^{i}/C_{n}^{n_2}$, where $C_n^m$ represents the number of combinations of $m$ items chosen from $n$ items. For example, when $n=6$, $n_1=2$, and $n_2=3$, the probability of one overlapping sample is $P_1=3/5$.
The distribution of the chance-overlap ratio can be estimated using a Gaussian function, especially when the sample size is large.
Consider the case where $n=100$, $n_1=20$, $n_2=30$, and the number of pre-selected truly associated items is $n_{\rm pick}=10$, the real proportion of associated items in the $n_1$ sample is $r_{\rm pick}=n_{\rm pick}/n_1=0.5$. The average proportion of items overlapping by chance is $r_{\rm chan}=\Sigma(P_i i/n_1)\simeq0.22$. Gaussian fitting of the distribution of the proportion of chance-overlap items indicates the most probable proportion of $r_{\rm chan,\,cen}=0.21$ with a variation of $r_{\rm chan,\,sig}=0.13$, which is consistent with $r_{\rm chan}$. The most probable observable overlap ratio is $r_{\rm obs}=(n_{\rm pick}+n_{\rm chan})/n_1\simeq0.61$, where $n_{\rm chan}$ is the most probable number of items overlapping by chance. 
If there is no pre-selected associated item, the probability of observing an overlap ratio of $r_{\rm obs}=0.61$ can be estimated from the distribution of chance-overlap ratios, which is $p_{\rm obs,\,chan}\simeq0.002$.
This hypothesis can be rejected at the 0.05 significance level, which is consistent with the scenario in which a subset of items is preselected as truly associated. Using the Gaussian fitting results for the proportion of overlapping items by chance, the proportion of associated items can be calculated as $r_{\rm asso}=(r_{\rm obs}-r_{\rm chan,\,cen})/(1-r_{\rm chan,\,cen})$, which is $0.49\pm0.19$, which is consistent with the actual proportion of associated items ($r_{\rm pick}$).
Therefore, by estimating the chance-overlap ratio, one can assess the likelihood of a chance overlap and estimate the proportion of truly associated samples.

\begin{figure}
\centering
\includegraphics[width=15cm, angle=0]{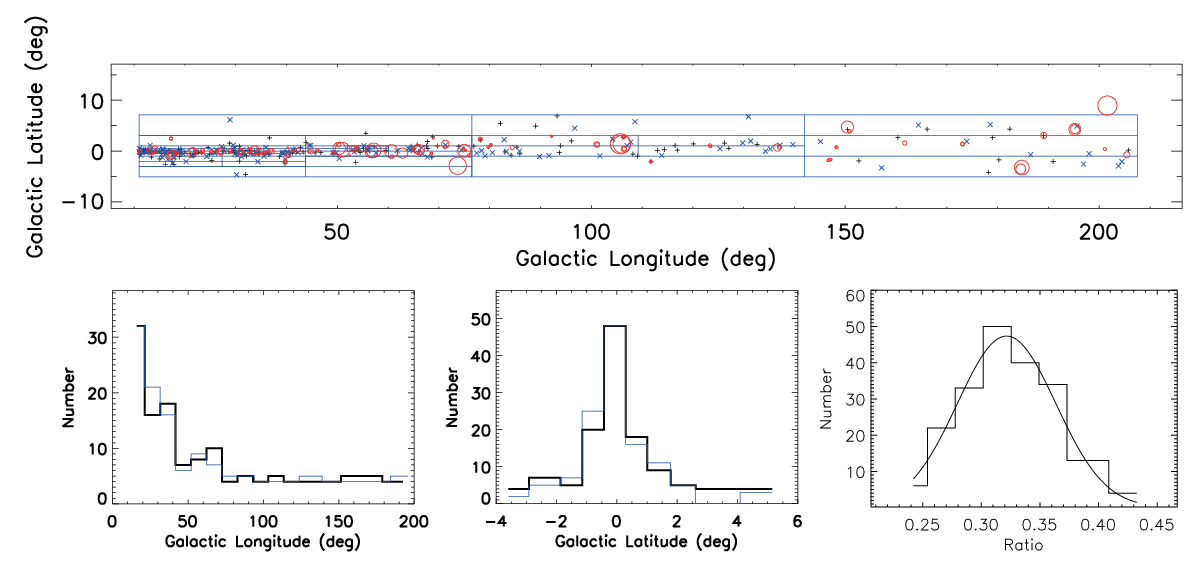}
\caption{Top: {\it l-b} map of the locations of LHAASO TeV gamma-ray sources (red circles) and known SNRs (excluding the PWN sample; black pluses). An example of randomly distributed SNRs (blue crosses) and the adaptive refined mesh grid used to generate them are also shown (see Section~\ref{sec:method} for details). 
The bottom left and middle panels show the {\it l} and {\it b} distributions of the SNRs (black) and of an example of randomly distributed SNRs (blue), respectively. 
The bottom right panel shows the distribution of the ratio of gamma-ray sources that overlap with randomly distributed SNRs. 
Each ratio value is calculated based on a set of randomly distributed SNRs.
The distribution of the ratio is fitted with a Gaussian function, and the the fitting result is shown by a black solid line. The Gaussian center and $1\sigma$ width of the random overlap ratio are 0.32 and 0.05, respectively.
The observed proportion of gamma-ray sources that overlap with the SNRs is $\sim$0.45.
}
\label{f:method}
\end{figure}

In practice, the chance-overlap ratio is evaluated as the proportion of sources in one class that overlap with sources in another class after the latter class has been randomized, as illustrated in Figure~\ref{f:method}.
For randomization, the sources are shuffled according to their Galactic longitude ($l$) and latitude ($b$) distributions. First, the $l$ and $b$ distributions of the target sources are fitted with Gaussian functions, and then a large number of random sources are generated according to these fitted distributions. Next, an adaptive refined mesh is constructed based on the target-source distribution, and the final random sources are selected so that each cell contains the same number of sources as the target sample. This adaptive mesh preserves the correlation between the Galactic longitude and latitude distributions of the target sources.
The size distribution of the sources is also constrained by the adaptive mesh and is randomized within each cell.
The chance-overlap ratio is repeatedly computed following this procedure until the resulting distribution is well described by a Gaussian, specifically, when the error in the fitted Gaussian center is an order of magnitude smaller than the fitted standard deviation ($\sigma$). 
In this way, the probability of a chance overlap between two source samples can be robustly evaluated. This method can thus verify the association of two source samples and estimate the proportion of truly associated sources. We refer to this approach as the Randomization-Adjusted Overlap Correlation (RAOC) method.
The same procedure can also be used to examine the correlation among the sources within a sample, by comparing the self-overlap ratio of an observed source sample with that evaluated by its corresponding randomized samples.
The self-overlap ratio is defined as the proportion of sources that overlap with one another within a sample.

\section{Results} \label{sec:result}
\subsection{Sample Association} \label{sec:asso}
\begin{table}
\tiny 
\begin{center}
\caption[]{Overlapping parameters between TeV gamma-ray sources and different objects}\label{tab:overpar}
 \begin{tabular}{C{25pt}C{30pt}C{30pt}C{23pt}C{25pt}C{40pt}C{43pt}C{33pt}C{40pt}}
  \hline\noalign{\smallskip}
Object$^{\rm (a)}$ &  Gamma-ray component$^{\rm (b)}$ & MC association$^{\rm (c)}$& $n_{\rm obj}$$^{\rm (d)}$ & $n_{\rm gamma}$$^{\rm (d)}$ & $r_{\rm gamma,\,obs}$$^{\rm (e)}$&$r_{\rm gamma,\,chan}$$^{\rm (f)}$& $r_{\rm obj,\,obs}$$^{\rm (g)}$& $r_{\rm obj,\,chan}$$^{\rm (h)}$\\
  \hline\noalign{\smallskip}
\HII\ region&All&All&3991&140&0.67&0.66$\pm$0.03&0.36&0.31$\pm$0.04\\
&WCDA&All&3978&65&0.69&0.71$\pm$0.04&0.28&0.17$\pm$0.03\\
&KM2A&All&3990&75&0.65&0.62$\pm$0.04&0.26&0.13$\pm$0.03\\
SNR&All&All&127&126&0.45&0.32$\pm$0.05&0.39&0.40$\pm$0.04\\
&&MC&108&125&0.39&0.31$\pm$0.04&0.41&0.43$\pm$0.05\\
&WCDA&All&126&60&0.50&0.34$\pm$0.05&0.33&0.25$\pm$0.05\\
&&MC&107&60&0.43&0.34$\pm$0.05&0.35&0.28$\pm$0.05\\
&KM2A&All&127&66&0.41&0.30$\pm$0.05&0.29&0.23$\pm$0.04\\
&&MC&108&65&0.35&0.30$\pm$0.05&0.31&0.24$\pm$0.05\\
PWN&All&All&18&119&0.22&0.03$\pm$0.03&0.72&0.29$\pm$0.09\\
&&MC&14&119&0.20&0.03$\pm$0.02&0.86&0.4$\pm$0.2\\
&WCDA&All&18&58&0.22&0.04$\pm$0.02&0.67&0.21$\pm$0.08\\
&&MC&14&58&0.21&0.04$\pm$0.02&0.79&0.2$\pm$0.2\\
&KM2A&All&18&61&0.21&0.04$\pm$0.02&0.61&0.11$\pm$0.06\\
&&MC&14&61&0.20&0.03$\pm$0.02&0.71&0.15$\pm$0.09\\
microquasar&All&All&14&142&0.03&0.0008$\pm$0.003&0.21&0.05$\pm$0.07\\
&WCDA&All&13&67&0.03&0.003$\pm$0.007&0.15&0.04$\pm$0.06\\
&KM2A&All&13&75&0.03&0.001$\pm$0.007&0.15&0.07$\pm$0.04\\
OB association&All&All&99&139&0.61&0.55$\pm$0.06&0.42&0.43$\pm$0.06\\
&WCDA&All&87&65&0.60&0.55$\pm$0.06&0.36&0.37$\pm$0.04\\
&KM2A&All&99&74&0.62&0.53$\pm$0.07&0.35&0.35$\pm$0.05\\
\\
  \noalign{\smallskip} \hline
\end{tabular}
\end{center}
\tablecomments{0.86\textwidth}{
\begin{itemize}
 \item[(a)] 
 SNRs and pure PWNe are introduced from the SNR- and PWN-molecular cloud (MC) association catalogs \citep[Tables 1 and 4 in][]{Zhou+2023}, respectively. The SNR sample does not include the pure PWN sample.
 \item[(b)] Component names of gamma-ray sources, corresponding to the detector names of the LHAASO for observing them \citep{Cao+2024}. 
The combined sample, treating each component as an independent source, is labeled 'All'. 
 \item[(c)] Association between objects and MCs; 'All' indicates the overall sample, and 'MC' indicates the sub-sample of objects that are associated with MCs. For the 'MC' sample, all possible associations are considered \citep[see][for details]{Zhou+2023}. 
 \item[(d)] Number of objects and gamma-ray sources; to match the coverages of different catalogs, sky cutting is performed, taking into account only objects and gamma-ray sources in areas where the coverages overlap.
 \item[(e)] Observed overlap proportion of gamma-ray sources with the objects.
 \item[(f)] Chance overlap proportion of gamma-ray sources, obtained from the proportion of gamma-ray sources overlapping with randomly distributed objects; see Section~\ref{sec:method} for details.
 \item[(g)] Observed overlap proportion of objects with gamma-ray sources.
 \item[(h)] Chance overlap proportion of objects, obtained from the proportion of objects overlapping with randomly distributed gamma-ray sources; see Section~\ref{sec:method} for details.
\end{itemize}
}
\end{table}

\begin{table}
\tiny 
\begin{center}
\caption[]{Association parameters between TeV gamma-ray sources and different objects}\label{tab:assopar}
 \begin{tabular}{C{50pt}C{42pt}C{35pt}cccc}
  \hline\noalign{\smallskip}
Object$^{\rm (a)}$ &  Gamma-ray component$^{\rm (a)}$ & MC association$^{\rm (a)}$& $p_{\rm gamma,\,chan}$$^{\rm (b)}$& $r_{\rm gamma,\,asso}$$^{\rm (c)}$& $p_{\rm obj,\,chan}$$^{\rm (d)}$& $r_{\rm obj,\,asso}$$^{\rm (e)}$\\
  \hline\noalign{\smallskip}
\HII\ region&All&All&0.28&0.04$\pm$0.09&0.10&0.07$\pm$0.06\\
&WCDA&All&0.77&$<$0.12&$2\times10^{-5}$&0.14$\pm$0.04\\
&KM2A&All&0.19&0.08$\pm$0.2&$3\times10^{-7}$&0.15$\pm$0.04\\
SNR&All&All&0.001&0.19$\pm$0.08&0.63&$<$0.05\\
&&MC&0.01&0.12$\pm$0.06&0.68&$<$0.05\\
&WCDA&All&0.0004&0.24$\pm$0.08&0.04&0.10$\pm$0.07\\
&&MC&0.03&0.14$\pm$0.08&0.08&0.09$\pm$0.07\\
&KM2A&All&0.02&0.15$\pm$0.08&0.07&0.08$\pm$0.06\\
&&MC&0.13&0.07$\pm$0.08&0.10&0.08$\pm$0.07\\
PWN&All&All&0.&0.20$\pm$0.04&$2\times10^{-7}$&0.6$\pm$0.2\\
&&MC&0.&0.18$\pm$0.03&$6\times10^{-5}$&0.8$\pm$0.5\\
&WCDA&All&0.&0.19$\pm$0.03&$2\times10^{-9}$&0.6$\pm$0.2\\
&&MC&0.&0.18$\pm$0.03&$2\times10^{-7}$&0.7$\pm$0.4\\
&KM2A&All&0.&0.18$\pm$0.03&0.&0.56$\pm$0.08\\
&&MC&0.&0.17$\pm$0.03&$2\times10^{-11}$&0.7$\pm$0.2\\
microquasar&All&All&0.&0.027$\pm$0.004&0.007&0.17$\pm$0.08\\
&WCDA&All&$5\times10^{-5}$&0.027$\pm$0.008&0.03&0.12$\pm$0.07\\
&KM2A&All&0.0002&0.025$\pm$0.008&0.02&0.09$\pm$0.05\\
OB association&All&All&0.11&0.1$\pm$0.2&0.55&$<$0.19\\
&WCDA&All&0.20&0.1$\pm$0.2&0.61&$<$0.05\\
&KM2A&All&0.07&0.2$\pm$0.2&0.51&$<$0.08\\
\\
  \noalign{\smallskip} \hline
\end{tabular}
\end{center}
\tablecomments{0.86\textwidth}{
\begin{itemize}
 \item[(a)] Samples are the same as those in Table~\ref{tab:overpar}.
 \item[(b)] Probability that the observed overlap in gamma-ray sources occurred by chance; see Section~\ref{sec:method} for details; extremely small numbers are rounded to zero.
 \item[(c)] Associated proportion of gamma-ray sources with the objects. The \HII\ regions exhibit a high self-overlap ratio, causing the chance-overlap ratio of the gamma-ray sources with them to be overestimated. Consequently, the best estimate of the associated proportion can be negative; in such cases, only an upper limit is provided.
 \item[(d)] Probability that the observed overlap in gamma-ray sources occurred by chance; see Section~\ref{sec:method} for details.
 \item[(e)] Associated proportion of objects with the gamma-ray sources. Because a large fraction of the WCDA and KM2A gamma-ray components overlaps with each other, the chance-overlap ratios of the objects with the combined gamma-ray source sample are overestimated. Therefore, the best estimate of the associated proportion can be negative, and only the upper limit is provided in such case.
\end{itemize}
}
\end{table}

The associations between TeV gamma-ray sources and various kinds of objects (i.e., pure PWNe, SNRs, \HII\ regions, microquasars, and OB associations) are examined based on the spatial correlation between their catalogs. If two kinds of sources are correlated, 
then a significant proportion of one kind of source should spatially overlap with another.
We assess the significance of the observed overlap by comparing it with randomized overlaps using the RAOC method (see Section~\ref{sec:method}).
In particular, subsamples of the SNR and PWN objects associated with MCs are studied individually. The overlapping and association parameters between TeV gamma-ray sources and different objects are listed in Tables~\ref{tab:overpar} and \ref{tab:assopar}, respectively.

The gamma-ray sources contain WCDA and KM2A components (see Section~\ref{sec:data}), and we examine the associations of these two components with different objects both individually and collectively. The two components are treated as independent sources. 
The self-overlap ratio of each component (WCDA and KM2A) is consistent with that of its corresponding randomized sample. 
However, the proportion of overlap between the WCDA and KM2A components is high. 
Approximately $\sim$80\% of the WCDA and KM2A sources overlap with each other, whereas the corresponding randomized sources overlap at a rate of only $\sim$30\%. This indicates that $\sim$70\% of the WCDA and KM2A sources share the same origin.
Splitting gamma-ray sources is equivalent to increasing the quantity of randomized gamma-ray sources. Consequently, the chance-overlap proportion of the objects with the combined gamma-ray source list is overestimated, and the corresponding association ratios underestimated. 
Therefore, the association results obtained with individual gamma-ray components should be adopted.

The correlations among the sources within each sample of other objects are also examined using the RAOC method (see Section~\ref{sec:method}).
The self-overlap ratios of the other kinds of objects are consistent with those of their corresponding randomized samples, except for SNRs and \HII\ regions.
SNRs have a self-overlap ratio of $\sim$0.03, however, the level of the chance-overlap ratio among SNRs is estimated to be 0.17$\pm$0.05 using the RAOC method (see Section~\ref{sec:method}).
Thus, the observed self-overlap ratio of SNRs is significantly lower than the corresponding chance-overlap ratio.
This suggests that the known SNRs exhibit a significant selection effect, with a substantial number of background SNRs remaining unidentified, amounting to $\sim$14\% of the known SNR sample \citep[e.g., a large and faint SNR candidate discovered to overlap with a known SNR;][]{Zhou+2024}. 
These unidentified SNRs likely follow a spatial distribution similar to that of the known SNRs.
This selection effect causes the chance-overlap ratio of the gamma-ray sources with the SNRs to be slightly underestimated, though not to a significant degree. 
The self-overlap ratio of the \HII\ regions ($\sim$0.55) is significantly higher than that of their corresponding randomized samples ($0.340\pm0.009$). This indicates that a substantial fraction of \HII\ regions are clustered together, with a clustering fraction of $\sim$0.3.
Consequently, the chance-overlap ratio of the gamma-ray sources with the \HII\ regions is overestimated, and the corresponding association ratio can be considered as a lower limit.

Overlap ratios are examined in two ways: the proportion of the gamma-ray sources that overlap with the objects, and the proportion of the objects that overlap with the gamma-ray sources. The former indicates whether the gamma-ray sources originate from the corresponding objects, while latter indicates whether the objects are gamma-ray bright.
These two approaches also differ in statistical significance, particularly for the overall gamma-ray sources, SNRs, and \HII\ regions, where the chance-overlap ratios may be over- or underestimated under one of the two approaches.
Association counts can be estimated by multiplying the association ratio by the number of gamma-ray sources or objects. The association numbers obtained by these two approaches are consistent in cases where the chance-overlap ratio is well determined. 

The origin of the gamma-ray sources is examined based on their associations with different kinds of objects.
Given that the probabilities ($p_{\rm gamma,\,chan}$ in Table~\ref{tab:assopar}) are below the 0.05 significance level, we can reject the hypothesis that the gamma-ray sources overlap with SNRs, PWNe, and microquasars by chance. Thus, a significant proportion of the gamma-ray sources originate from SNRs, PWNe, and microquasars.
The proportion of gamma-ray sources associated with SNRs is estimated to be 0.19$\pm$0.08; even when the selection effect in the SNR sample is considered, the proportion remains $\sim$0.17. 
The association proportion with SNRs is higher for WCDA gamma-ray sources (0.24$\pm$0.08) and lower for KM2A gamma-ray sources (0.15$\pm$0.08).
The proportion of gamma-ray sources associated with PWNe is estimated as 0.20$\pm$0.04, with similar values for the WCDA and KM2A components. 
Although the pure PWNe are excluded from the SNR sample, some SNRs in our selection still contain PWN sources (e.g., composite-type SNRs). 
We also examine the subsample of shell-type SNRs and find that the proportion of the gamma-ray sources associated with them is $\sim$0.1.
For the microquasars, a small but significant proportion of gamma-ray sources is associated with them (0.027$\pm$0.004), which is similar for the WCDA and KM2A components.

The high probability of chance overlap for the gamma-ray sources with the \HII\ regions is overestimated due to the significant self-overlap among the \HII\ regions themselves. After accounting for this effect, the KM2A gamma-ray sources are likely associated with \HII\ regions. 
In contrast, the chance-overlap probability of the gamma-ray sources with the OB associations is high, suggesting a limited contribution of OB associations to the observed gamma-ray emission. 

The gamma-ray emission capability of different kinds of objects is examined based on their overlap proportion with gamma-ray sources.
Because a large fraction of the WCDA and KM2A gamma-ray components overlap with each other, the chance-overlap ratios of the objects with the combined gamma-ray source sample are overestimated. Hence, the independent results for the WCDA and KM2A components are more reliable. 
As indicated by the chance-overlap probability, PWNe, SNRs, microquasars, and \HII\ regions are associated with the TeV gamma-ray sources, whereas OB associations are not. 
The proportion of PWNe associated with the gamma-ray sources is $\sim$60\%, which is similar for both the WCDA and KM2A components. 
SNRs tend to be associated with the WCDA sources at a rate of $\sim$10\%.
The proportion of microquasars associated with the gamma-ray sources is $\sim$12\% for the WCDA component and $\sim$9\% for the KM2A component.
For the \HII\ region sample, the associated proportion is $\sim$15\%, again similar for both gamma-ray components. 
We note that the number of the \HII\ regions associated with the gamma-ray sources is relatively large due to their high self-overlap ratio.

We also separately examine subsamples of SNRs and PWNe that are associated with MCs. 
Because they have relatively small sample sizes, the association ratios of the gamma-ray sources with these objects are lower than those for the overall SNR and PWN samples, respectively. 
The association ratios of the SNR-MC subsample with the gamma-ray sources are comparable to those of the overall SNR sample. 
Here, we do not consider gamma-ray sources that are associated with MCs passively illuminated by nearby SNRs but do not spatially overlap with the SNRs \citep[e.g.,][]{MitchellCelli2024}. 
Notably, the gamma-ray-bright proportion of the the shell-type SNR subsample associated with MCs is enhanced.
The association ratios of the PWN-MC subsample with the gamma-ray sources are slightly higher than those of the overall PWN sample. 
These results suggest that association with MCs increases the likelihood for SNRs and PWNe to be gamma-ray bright.

\subsection{Gamma-ray Bright MCs around PWNe} \label{sec:pwnmc}
\begin{figure}
\centering
\includegraphics[width=13cm, angle=0]{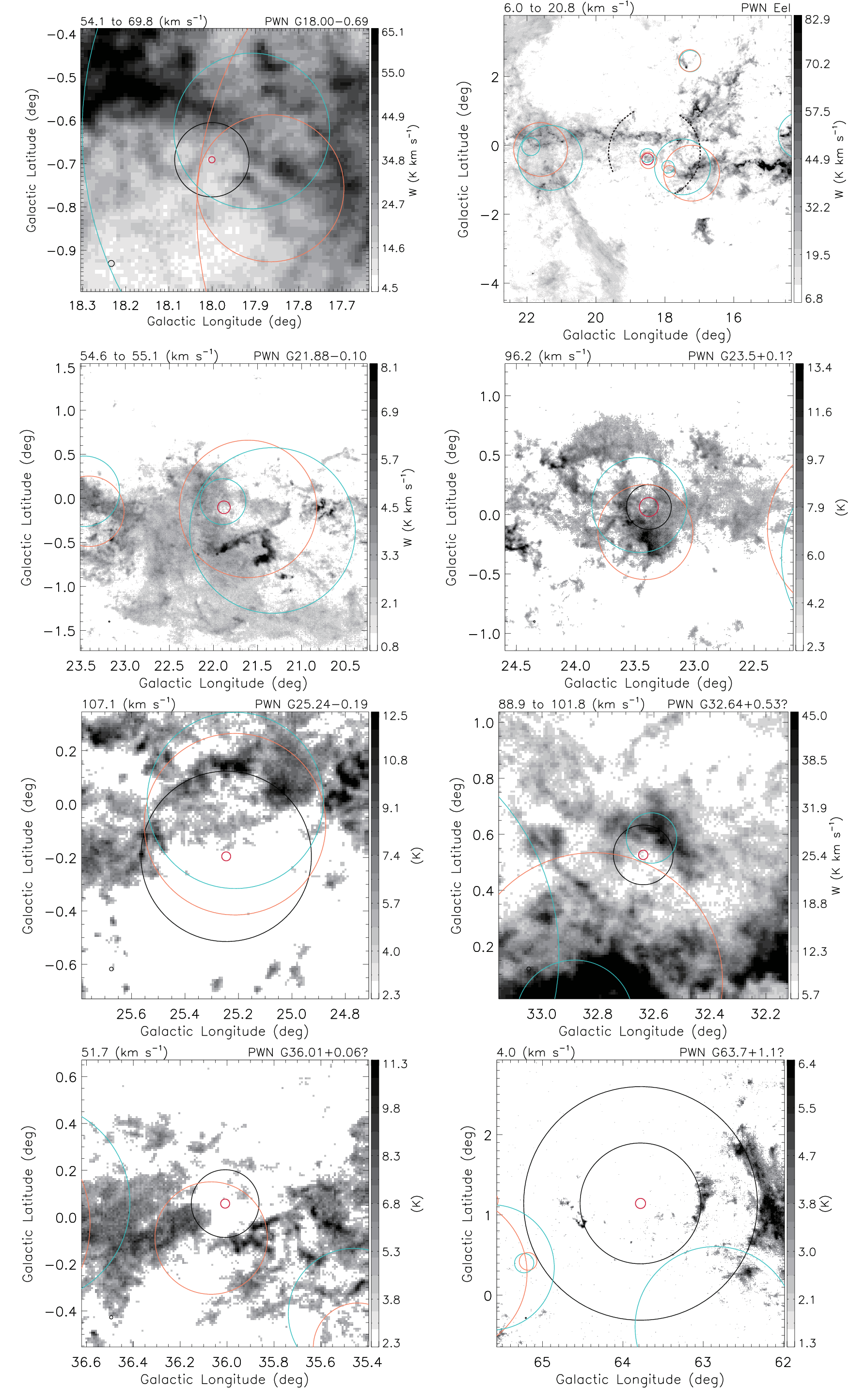}
\caption{\twCO\ (J=1--0) intensity maps of the candidates of the associated MCs around PWNe G18.00$-$0.69, Eel, G21.88$-$0.10, G23.5$+$0.1, G25.24$-$0.19, G32.64$+$0.53, G36.01$+$0.06, and G63.7$+$1.1 \citep[see][for details]{Zhou+2023}.
The intensities of the \twCO\ emission are at least $3\sigma$. 
The extents of the PWNe are indicated by red circles.
The surrounding shell-like structures are outlined by black solid circles, except for PWN Eel, which is outlined by black dotted lines. Orange and cyan solid circles denote the extents of WCDA and KM2A gamma-ray sources, respectively.
The beam is represented by a black circle in the lower left corner.
}
\label{f:pwnli1}
\end{figure}
\begin{figure}
\centering
\includegraphics[width=15cm, angle=0]{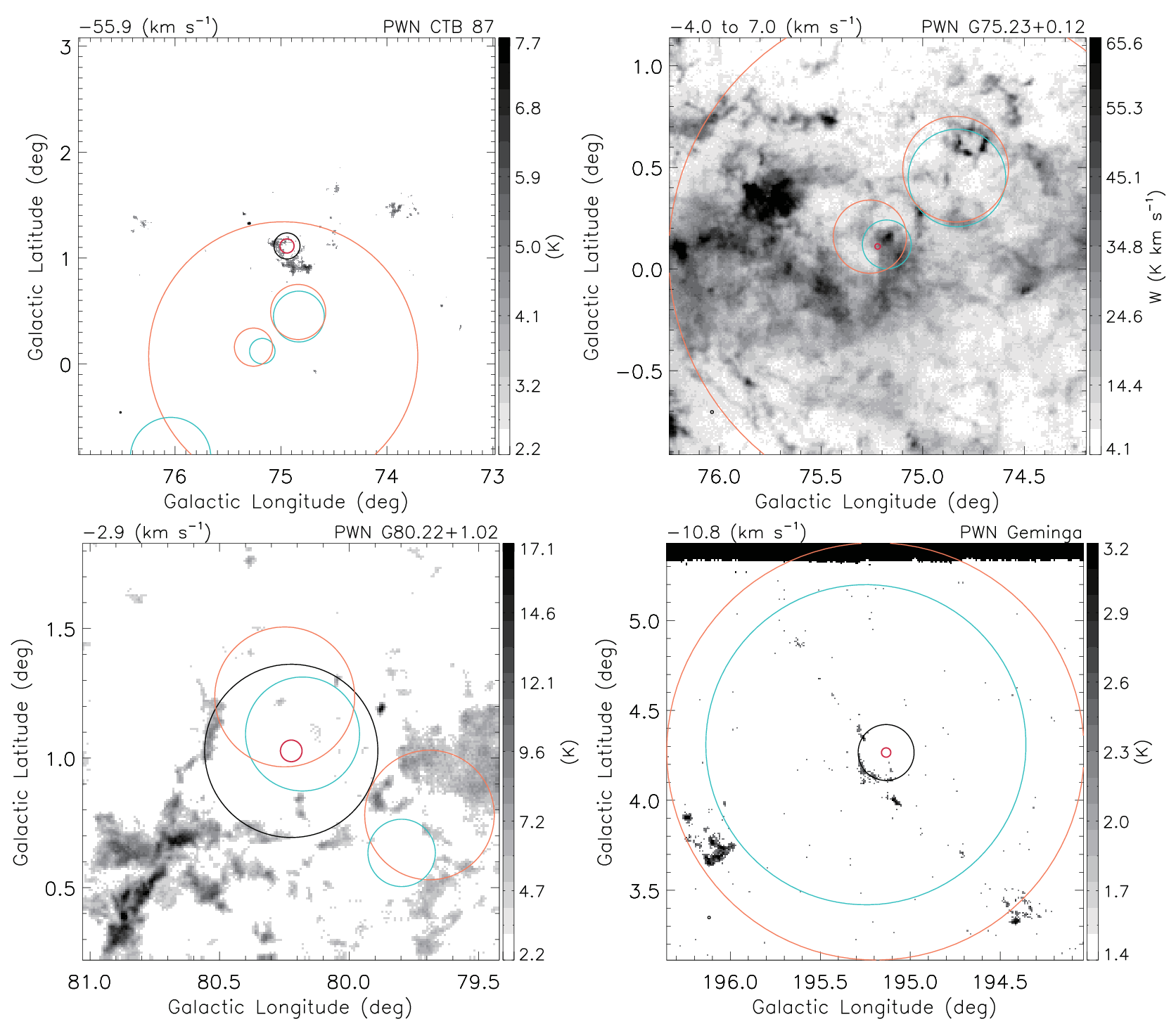}
\caption{Same as Figure~\ref{f:pwnli1}, but for PWNe CTB~87, G75.23$+$0.12, G80.22$+$1.02, and Geminga.
}
\label{f:pwnli2}
\end{figure}

The gamma-ray bright proportion of MC-associated PWNe appears to be higher. To verify this, we further compare the positions of the gamma-ray sources with the distributions of MCs associated with PWNe.
Figures~\ref{f:pwnli1} and \ref{f:pwnli2} illustrate the distributions of the gamma-ray sources and candidate-associated MCs around each PWN. 
PWN-MC association candidates are introduced from \cite{Zhou+2023}.
In general, the gamma-ray sources that overlap with PWNe are offset toward the associated MCs. 
In the case of the PWN Eel, additional gamma-ray sources overlap with the dense part of the associated molecular shell. 
These results indicate that gamma-ray emission from PWNe is related to the surrounding MCs. 
We note that the gamma-ray sources overlapping PWN G80.22+1.02 are not offset toward the candidate-associated MC, and the gamma-ray sources overlapping the Crab PWN align closely with the PWN itself.
It is also notable that PWN G63.7$+$1.1 does not overlap with any gamma-ray source; nevertheless, it has been detected in neutrino emission, which supports the presence of a proton-proton (p-p) interaction process, with dense molecular gas providing abundance target protons \citep{Ji+2025}.
Further dedicated studies of these PWN-MC association candidates are needed to examine the properties of the corresponding gamma-ray emission.

The tendency of MC-associated PWNe to be bright in gamma-rays suggests the presence of high-energy protons that interact with the dense gas in MCs, producing gamma-rays via p-p collisions. This finding supports the possibility that PWNe may contribute to both leptonic and hadronic CR populations. Alternatively, it is also possible that the evolution of PWNe is influenced by associated MCs, leading to changes in their gamma-ray emission \citep[e.g., a scenario similar to that proposed for HESS J1809$-$193;][]{SunJ+2025}.

%
%
%

\section{Summary}\label{sec:sum}
We examine the associations between LHAASO TeV gamma-ray sources and different kinds of objects, such as PWNe, SNRs, \HII\ regions, microquasars, and OB associations.
The pure PWN sample is excluded from the SNR sample.
The subsamples of the SNRs and PWNe that are associated with MCs are specifically examined. 
We propose the RAOC method to evaluate the probability of associations among sources from multiple catalogs and to estimate the proportion of associated sources.
We emphasize that the RAOC method can serve as a universal approach for assessing correlations between two samples or within an individual sample.

LHAASO TeV gamma-ray sources exhibit a statistically significant overlap with SNRs, PWNe, and microquasars, suggesting that these objects are important contributors to gamma-ray emission. We estimate the association proportions of the gamma-ray sources to be 0.19$\pm$0.08 for SNRs, 0.20$\pm$0.04 for PWNe, and 0.027$\pm$0.008 for microquasars.
The proportion of the low energy WCDA gamma-ray sources that are associated with the SNRs is higher than that of the high energy KM2A gamma-ray sources.
We also find that the proportion of the gamma-ray sources associated with the subsample of shell-type SNRs is around 0.1.
The proportions of the gamma-ray sources that are associated with the PWNe and microquasars are similar for WCDA and KM2A.
Due to the high self-overlap ratio of \HII\ regions, the probability of the gamma-ray sources coincidentally overlapping with them is overestimated. The KM2A gamma-ray sources are probably associated with the \HII\ regions.
The probability that the gamma-ray sources coincidentally overlap with the OB associations is high, indicating their limited contributions to gamma-ray emissions.

We also examine the gamma-ray emission capabilities of different kinds of objects. Approximately 60\% of PWNe are gamma-ray bright in both the WCDA and KM2A components. 
The SNRs tend to be associated with low-energy gamma-ray sources (the WCDA component), with a proportion of about 10\%.
The association proportions of the microquasars are $\sim$12\% for the WCDA gamma-ray component and $\sim$9\% for the KM2A component.
These findings support the role of PWNe, SNRs, and microquasars as CR accelerators. 
The \HII\ regions are also associated with the gamma-ray sources, which demonstrates their ability to emit gamma-rays.
However, the proportion of gamma-ray-bright \HII\ regions cannot be determined well due to their large self-overlap ratio.

The self-correlation of different samples is also considered. Most source samples show no significant correlation within their own members, except for SNRs, \HII\ regions, and the WCDA and KM2A components of gamma-ray sources.
The results indicate that the majority of the WCDA and KM2A components of the gamma-ray sources have the same origin, with a proportion of about 70\%. 
We also find that known SNRs exhibit selection effects, resulting in many background SNRs remaining unidentified, amounting to about 14\% of the known SNR sample.
Possible reasons for this include the difficulty in detecting large and faint SNRs, as well as confusion between overlapping distinct SNRs.
For \HII\ regions, about 30\% of them are grouped together and are probably part of larger star-forming regions.

The subsamples of the SNRs and PWNe that are associated with MCs are examined independently.
The gamma-ray association proportions for the overall SNR sample and the SNR-MC subsample are comparable, which may be attributed to the small proportion values and limited sample sizes.
Nevertheless, the gamma-ray bright proportion for shell-type SNRs associated with MCs is enhanced.
The PWNe associated with MCs also exhibit higher association ratios with the gamma-ray sources, suggesting that these PWNe are more likely to be gamma-ray bright. 
We further compare the positions of gamma-ray sources to the distributions of MCs associated with PWNe. The results show that the gamma-ray sources overlapping PWNe are generally offset toward the related MCs, suggesting that MCs play a role in producing gamma-ray emission.
In-depth analysis of gamma-ray observations is needed to uncover the underlying physical mechanisms.

Further study on the relationship between gamma-ray sources and other kinds of objects can provide more insight into the origin of gamma-ray emission.

\begin{acknowledgements}
We are grateful to all the members of the MWISP working group, particularly the staff members at PMO-13.7 m telescope, for their long-term support. 
This work is partially supported by the National SKA Program of China (No. 2025SKA0140100).
This research made use of the data from the MWISP project, which is a multi-line survey in 12CO/13CO/C18O along the northern Galactic plane with PMO-13.7 m telescope. 
MWISP was sponsored by National Key R\&D Program of China with grants 2023YFA1608000 \& 2017YFA0402701 and by CAS Key Research Program of Frontier Sciences with grant QYZDJ-SSW-SLH047.
\end{acknowledgements}

 

\bibliographystyle{raa}

\label{lastpage}
\end{document}